\documentclass[twoside]{article}
\def\R{{I\!\! R}}
\def\N{{I\!\! N}}

\def\C{~\hbox{\vrule width 0.6pt height 6pt depth 0pt \hskip -3.5pt}C}
\def\kasten{$~~\mbox{\hfil\vrule height6pt width5pt depth-1pt}$ }

\newtheorem{theorem}{Theorem}[section]
\newtheorem{axioms}[theorem]{Axioms}
\newtheorem{proposition}[theorem]{Proposition}
\newtheorem{corrollary}[theorem]{Corollary}
\newtheorem{definition}[theorem]{Definition}
\newtheorem{lemma}[theorem]{Lemma}
\newtheorem{remark}[theorem]{Remark}
\newtheorem{condition}[theorem]{Condition}
\begin{document}
\pagestyle{myheadings} \markboth{ S. Albeverio and H.Gottschalk}{
Scattering theory for QFTs with Indef. Metric}
\thispagestyle{empty}
\begin{flushleft}
 {\Large \bf Scattering Theory for Quantum \linebreak
Fields with Indefinite Metric}

\

\noindent Sergio Albeverio and Hanno Gottschalk\\

Institut f\"ur angewandte Mathematik, \\ Rheinische
Friedrich-Wilhelms-Universit\"at Bonn,\\ Wegelerstr. 6, D-53115
Bonn, Germany\\ e-mail: albeverio@uni-bonn.de  and
gottscha@wiener.iam.uni-bonn.de \markboth{S. Albeverio and H.
Gottschalk}{Scattering Theory for QFTs with Indef. Metric}
\end{flushleft}
\begin{abstract}

In this work, we discuss the scattering theory of local,
relativistic quantum fields with indefinite metric. Since the
results of Haag--Ruelle theory do not carry over to the case of
indefinite metric \cite{AGW6}, we propose an axiomatic framework
for the construction of in- and out- states, such that the LSZ
asymptotic condition can be derived from the assumptions. The
central mathematical object for this construction is the
collection of mixed vacuum expectation values of local, in- and
out- fields, called the ``form factor functional'', which is
required to fulfill a Hilbert space structure condition. Given a
scattering matrix with polynomial transfer functions, we then
construct interpolating, local, relativistic quantum fields with
indefinite metric, which fit into the given scattering framework.
\end{abstract}
{\small \noindent {\bf Keywords}: {\it QFT with indefinite metric,
scattering theory, interpolating fields.}}
\section{Introduction}
The Wightman framework of local, relativistic quantum field theory
(QFT) turned out to be too narrow for theoretical physicists, who
were interested in handling situations involving in particular
gauge fields (like in quantum electrodynamics). For several
reasons which are intimately connected with the needs of the
standard procedure of the perturbative calculation of the
scattering matrix (for a detailed discussion, see \cite{Str}), the
concept of QFT with indefinite metric was introduced, where a
probability interpretation is possible only on Hilbert subspaces
singled out by a gauge condition in the sense of  Gupta \cite{Gu}
and Bleuler \cite{Bl}. On the other hand, "ghosts", which are
quantum fields with the "wrong" connection of spin and statistics,
entered the physical scene in connection with the
Fa\-de\-ev--Popov determinant in perturbation theory \cite{FP}. As
a consequence of Pauli's spin and statistics theorem, such quantum
fields can not be  realized  on a state space with positive
metric.

Mathematical foundations for QFT with indefinite metric were laid
by several authors, among them Scheibe \cite{Sc}, Yngvason
\cite{Yn}, Araki \cite{Ar}, Morchio and Strocchi \cite{MoSt},
Mintchev \cite{Min} and more recently by G. Hoffmann, see e.g.
\cite{Hoff2}. The results obtainable from the axioms of indefinite
metric QFT in many aspects are less strong than the axiomatic
results of positive metric QFT. As the richness of the axiomatic
results can be seen as a measure for the difficulty to construct
theories which fulfill such axioms \cite{SW}, the construction of
indefinite metric quantum fields can be expected to be simpler
than that of positive metric QFTs.

Up to now, however, the linkage between these mathematical
foundations and scattering theory, which in the day to day use of
physicists is based on the LSZ reduction formalism \cite{LSZ},
remained open, since the only available axiomatic scattering
theory (Haag-Ruelle theory\cite{Ha,Ru,He}) heavily relies on the
positivity of Wightman functions. One can even give explicit
counter examples of local, relativistic QFTs with indefinite
metric \cite{AGW6,BGL}, such that the LSZ asymptotic condition
fails and Haag-Ruelle like scattering amplitudes diverge
polynomially in time \cite{AGW6}.

A scattering theory for QFTs with indefinite metric which fits
well to the LSZ formalism and the mathematically rigorous
construction of models of indefinite metric quantum fields (in
arbitrary space-time dimension) with nontrivial scattering
behavior are the topic of this work, which is organized as
follows:

In the second section (and Appendix A) we set up the frame of QFT
with indefinite metric and recall some GNS-like results on the
representation of $*$-algebras on state spaces with indefinite
inner product.

In Section 3 we introduce a set of conditions which is taylored
just in the way to imply the LSZ asymptotic condition. The main
mathematical object is the collection of mixed expectation values
of incoming, local and outgoing fields, called "form factor
functional", which is required to fulfill a Hilbert space
structure condition (HSSC), cf. \cite{Hoff2,MoSt}. The existence
of the form factor functional can be understood as a restriction
of the strength of mass-shell singularities in energy-momentum
space which rules out the counter examples in \cite{AGW6}.

In Section 4 we construct a class of QFTs with indefinite metric
and nontrivial scattering behaviour fitting into the frame of
Section 3. The main ingredient of this section is a sequence of
local, relativistic truncated Wightman functions called the
'structure functions', which have been introduced and studied in
[1--5,8,17,18,25]. The non trivial scattering behaviour of the
structure functions has been observed in \cite{AGW4,Go2,Jo}. The
class of such QFTs is rich enough to interpolate essentially all
scattering matrices with polynomial transfer
functions\footnote{Schneider, Baumg\"artel and Wollenberg
constructed a class of weakly local interpolating QFTs with
positive metric\cite{BW,Sn}. These fields however can not be local
and are not related to the models we study here.}. Some technical
proofs can be found in Appendix B.

Section 5 is a supplement to Section 4, in which we discuss the
approximation of any set of measurement data for energies below a
maximal experimental energy $E_{\rm max}$ up to an experimental
accuracy given by an error tolerance $\epsilon>0$ with models in
the class of Section 4.

\section{Quantum fields with indefinite metric}
In this section we introduce our notation and we collect some
known facts about quantum field theories with indefinite metric
following \cite{AGW3,Hoff1,Hoff2,MoSt}.

In order to keep notations simple we study Bosonic, chargeless
QFTs\footnote{All results of this article can be generalized to
fields with arbitrary parameters, cf. \cite{Go2}.} over a $d$
dimensional Minkowski space-time $(\R^d,\cdot )$ where $ x\cdot
y=x^0y^0-{\bf x} \cdot {\bf y}$ for $x=(x^0,{\bf
x})=(x^0,x^1,\ldots,x^{d-1}),y=(y^0,{\bf
y})=(y^0,y^1,\ldots,y^{d-1})\in \R^d$. For $x\cdot x$ we will
frequently write $x^2$. The collection of all $k\in \R^d$ with
$k^2>m^2\geq 0$ and $k^0>0$ ($k^0<0$) is called the forward
(backward) mass-cone of mass $m$ and is denoted by the symbol
$V_m^+$ ($V_m^-$). By $\bar V_m^\pm$ we denote the closure of
$V^\pm_m$. The (topological) boundary of $V_m^+$ ($V_m^-$) is
called the forward (backward) mass shell. By ${\cal L}$ we denote
the full Lorentz group and by $\tilde {\cal P}^\uparrow_+$ the
(covering group of the) orthochronous, proper Poincar\'e group.

${\cal S}_n$ stands for the the complex valued Schwartz functions
over $\R^{dn}$and we set ${\cal S}_0=\C$. The topology on the
spaces ${\cal S}_n$ is induced by the Schwartz norms
 \begin{equation}
\label{2.1eqa} \| f\|_{K,L}=\sup_{ x_1,\ldots,x_n\in \R^{d}\atop
0\leq |\beta_1|,\ldots,|\beta_n|\leq K}\left|
\prod_{l=1}^n(1+|x_l|^2)^{L/2} D^{\beta_1\cdots \beta_n}
f(x_1,\ldots,x_n)\right|
\end{equation}
where $K,L\in\N$, $\beta_l=(\beta_l^1,\ldots,\beta_l^{d-1})\in
\N_0^{d}, l=1\ldots,n,$ are multi indices with $|\beta_l|=
\sum_{j=0}^{d-1}\beta_l^j$,
$D^{\beta_1\ldots\beta_n}=\prod_{l=1}^n(\partial^{|\beta_l|}/\partial
x_l^{\beta_l})$.

The canonical representation $\alpha$ of $\tilde {\cal
P}^\uparrow_+$ on ${\cal S}_n$ is given by
\begin{equation}
\label{2.7beqa}
\alpha_{\{\Lambda,a\}}f(x_1,\ldots,x_n)=f(\Lambda^{-1}(x_1-a),\ldots,\Lambda^{-1}(x_n-a)
)
\end{equation}
$\forall \{\Lambda,a\}\in \tilde P^\uparrow_+,f\in S_n.$

We normalise the Fourier transform ${\cal F}:{\cal S}_n\to {\cal
S}_n$ as follows
\begin{equation}
\label{2.8eqa}
 {\cal F}f(k_1,\ldots,k_n)= (2\pi)^{-dn/2} \int_{\R^{dn}}\! e^{ -i ( x_1\cdot k_1+\cdots+ x_n\cdot k_n)} f(x_1,\ldots,x_n) ~dx_1\cdots dx_n
\end{equation}
$\forall f\in {\cal S}_n.$ Frequently we will also use the
notation $\hat f$ instead of ${\cal F} f$. For the inverse Fourier
transform of $f$ we write $\bar {\cal F}f$.

Let $\underline{\cal S}$ be the Borchers' algebra over ${\cal
S}_1$, namely $\underline{\cal S}=\bigoplus_{n=0}^\infty {\cal
S}_n$.  $\underline{f}\in \underline{\cal S}$ can be written in
the form $\underline{f}=(f_0,f_1,\ldots,f_j,0,\ldots,0,\ldots)$
with $f_0\in\C$ and $f_n\in{\cal S}_n$,  $j\in\N$.

The addition and multiplication on $\underline{\cal S}$ are
defined as follows:
\begin{equation}
\label{2.14eqa}
\underline{f}+\underline{h}=(f_0+h_0,f_1+h_1,\ldots)
\end{equation}
and
\begin{eqnarray}
\label{2.15eqa}
\underline{f}\otimes\underline{h}&=&((\underline{f}\otimes
\underline{h})_0,(\underline{f}\otimes\underline{h})_1,\ldots)\nonumber
\\ (\underline{f}\otimes\underline{h})_n&=& \sum_{j,l=0\atop
j+l=n}^\infty f_j\otimes h_l ~~\mbox{for}~n\in \N_0
\end{eqnarray}
The involution $*$, the Fourier transform $\underline{\cal F}$ and
the representation $\underline{\alpha }$ of $\tilde {\cal
P}^\uparrow_+$ on $\underline{\cal S}$ are defined through

\begin{eqnarray}
\label{2.16eqa}
\underline{f}^*&=&(f_0^*,f_1^*,f_2^*\ldots)\nonumber\\
\underline{\cal F}\underline{f}&=&(f_0,{\cal F}f_1,{\cal
F}f_2,\ldots)\\
\underline{\alpha}_{\{\Lambda,a\}}\underline{f}&=&(f_0,\alpha_{\{\Lambda,a\}}f_1,\alpha_{\{\Lambda,a\}}f_2,\ldots)\nonumber
\end{eqnarray}
where $f_n^*(x_1,\ldots,x_n)=\overline{f_n(x_n,\ldots,x_1)}$.

 We
endow $\underline{\cal S}$ with the strongest topology, such that
the relative topology of ${\cal S}_n$ in $ \underline{\cal S}$ is
the Schwartz topology (direct sum topology). Let $\underline{\cal
S}'=\underline{\cal S}'(\R^d,\C)$ be the topological dual space of
$\underline{\cal S}$. Then $\underline{R}\in\underline{\cal S}'$
is of the form $\underline{R}=(R_0,R_1,R_2,\ldots)$ with $R_0\in
\C, R_n\in {\cal }S_n', n\in\N$. Furthermore, any such sequence
defines uniquely an element of $\underline{\cal S}'$. As in the
case of $\underline{\cal S}$, the involution, Fourier transform
and representation of $\tilde {\cal P}^\uparrow_+$ are on
$\underline{S}'$ are defined by the corresponding actions on the
components ${\cal S}'_n$ of $\underline{\cal S}'$.

Elements of $\underline{\cal S}'$ are also called Wightman
functionals. The tempered distributions $W_n\in {\cal S}_n'$
associated to a Wightman functional
$\underline{W}\in\underline{\cal S}'$ are also called (n-point)
Wightman functions.

Next we introduce the modified Wightman axioms of Morchio and
Strocchi for QFTs in indefinite metric.

\begin{axioms}
\label{2.1ax} \rm A1) Temperedness and normalization:
$\underline{W}\in\underline{\cal S}'$ and $W_0=1$.

\noindent A2) Poincar\'e invariance:
$\underline{\alpha}_{\{\Lambda,a\}}\underline{W}=\underline{W}~
\forall \{\Lambda,a\}\in \tilde {\cal P}^\uparrow_+$.

\noindent A3) Spectral property: Let $ I_{\mbox{sp}}$ be the left
ideal in $\underline{\cal S}$ generated by elements of the form
$(0,\ldots,0,f_n,0\ldots)$ with $\mbox{supp}\hat f_n\subseteq \{
(k_1,\ldots,k_n)\in \R^{dn}:\sum_{l=1}^nk_l\not\in \bar V_0^+\}$.

Then $I_{\mbox{sp}}\subseteq \mbox{kernel}~\underline{W}$.

\noindent A4) Locality: Let $ I_{\mbox{loc}}$ be the two-sided
ideal in $ \underline{ \cal S}$ generated by elements of the form
$(0,0,[f_1,h_1],0,\ldots)$ with $\mbox{supp}~f_1$ and $
\mbox{supp}~h_1$ space-like separated.

Then $I_{\mbox{loc}}\subseteq \mbox{kernel}~\underline{W}$.

\noindent A5) Hilbert space structure condition (HSSC): There
exists a Hilbert seminorm $\underline{p}$ on $\underline{\cal S}$
s.t. $\left|\underline{W}(\underline{f}^*\otimes
\underline{g})\right|\leq
\underline{p}(\underline{f})\underline{p}(\underline{g}) \forall
\underline{f},\underline{g} \in \underline{S}$.

\noindent A6) Cluster Property: $\lim_{t\to
\infty}\underline{W}(\underline{f}\otimes \underline{\alpha
}_{\{1,ta\}}\underline {g}) =
\underline{W}(\underline{f})~\underline{W}(\underline{h})$
$\forall \underline{f},\underline{g}\in \underline{S}, a\in \R^{d}
$ space like (i.e. $a^2<0$).

\noindent A7) Hermiticity: $\underline{W}^*=\underline{W}$.
\end{axioms}

All these axioms can be equivalently expressed in terms of
Wightman functions in the usual way, cf. \cite{Dopl,MoSt,SW}.

The significance of the Axioms \ref{2.1ax} can be seen from the
following GNS-like construction:

A metric operator $\eta:{\cal H}\to{\cal H}$ by definition is a
self adjoint operator on the separable Hilbert space $({\cal
H},(.,.))$ with $\eta^2=1$. Let  ${\cal D}$ be a dense and linear
subspace. We denote the set of (possibly unbounded) Hilbert space
operators $A:{\cal D}\to{\cal D}$ with (restricted) $\eta$-adjoint
$A^{[*]}=\eta A^*\eta|_{\cal D}:{\cal D}\to{\cal D}$ with ${\sf
O}_\eta ({\cal D})$. Clearly, ${\sf O}_\eta ({\cal D})$ is an
unital algebra with involution $[*]$. The canonical topology on
${\sf O}_\eta ({\cal D})$ is generated by the seminorms
$A\to|(\Psi_1,\eta A \Psi_2)|,~\Psi_1,\Psi_2\in {\cal D}$. We then
have the following theorem:

\begin{theorem}
\label{2.1theo} Let $\underline{W}\in\underline{\cal S}$ be a
Wightman functional which fulfills the Axioms \ref{2.1ax}. Then

\noindent (i) There is a Hilbert space $({\cal H},(.,.))$ with a
distinguished normalized vector $\Psi_0\in {\cal H}$ called the
vacuum, a metric operator $\eta$ with $\eta\Psi_0=\Psi_0$ inducing
a nondegenerate inner product $\langle.,.\rangle=(.,\eta .)$ and a
continuous $*$-algebra representation $\phi:\underline{\cal S}\to
{\sf O}_\eta ({\cal D})$ with ${\cal D}=\phi(\underline{S})\Psi_0$
which is connected to the Wightman functional $\underline{W}$ via
$\underline{W}(\underline{f})=\langle
\Psi_0,\phi(\underline{f})\Psi_0\rangle\forall \underline{f}\in
\underline{\cal S}$.

\noindent (ii) There is a $\eta$-unitary continuous representation
${\sf U}: \tilde {\cal P}_+^\uparrow \to{\sf O}_\eta ({\cal D})$
($\, {\sf U}^{[*]}={\sf U}^{-1}$) such that ${\sf U}( \Lambda,a)
\phi (\underline{f}){\sf
U}(\Lambda,a)^{-1}=\phi(\underline{\alpha}_{\{\Lambda,a\}}^{-1}\underline{f})~\forall
\underline{f}\in\underline{\cal S}, \{\Lambda,a\}\in \tilde {\cal
P}_+^\uparrow$ and $\Psi_0$ is invariant under the action of ${\sf
U}$.

\noindent (iii) $\phi$ fulfills the spectral condition
$\phi(I_{\mbox{\rm sp}})\Omega=0$.

\noindent (iv) $\phi$ is a local representation in the sense that
$I_{\mbox{\rm loc}} \subseteq \mbox{\rm kernel}~\phi$.

\noindent (v) For $\Psi_1,\Psi_2\in{\cal D}$ and $a\in \R^d$ space
like, we get $\lim_{t\to\infty}\langle\Psi_1,{\sf
U}(1,ta)\Psi_2\rangle=\langle\Psi_1,\Psi_0\rangle\langle\Psi_0,\Psi_2\rangle$.

A quadruple $(({\cal H},\langle.,.\rangle,\Psi_0),\eta,{\sf
U},\phi)$ is called a local relativistic QFT with indefinite
metric.

Conversely, let $(({\cal H},\langle.,.\rangle,\Psi_0),\eta,{\sf
U},\phi)$ be a local relativistic QFT with indefinite metric. Then
$ \underline{W}(\underline{f})=\langle
\Psi_0,\phi(\underline{f})\Psi_0\rangle  ~ \forall
\underline{f}\in\underline{\cal S}$ defines a Wightman functional
$\underline{W}\in\underline{\cal S}'$ which fulfills the Axioms
\ref{2.1ax}.
\end{theorem}

\noindent {\it Proof.} See \cite{Bo2,MoSt}. For the fact that we
can chose the metric operator in such a way that
$\eta\Psi_0=\Psi_0$, cf. \cite{Hoff2}. Item (v) is just a
rephrasing of the cluster property (A6). \kasten

It should be mentioned that the pair
$(\underline{W},\underline{p})$ uniquely determines the associated
QFT with indefinite metric, but it is believed that in general the
Wightman functional $\underline{W}$ admits non equivalent
representations as the vacuum expectation value of a QFT with
indefinite metric depending on the choice of $\underline{p}$, cf.
\cite{Ar} for a related situation. See however \cite{Hoff2} for
sufficient conditions s.t. only $\underline{W}$ determines the
(maximal) Hilbert space structure.

We want to study sufficient topological conditions on the Wightman
functionals which imply the HSSC and therefore the existence of
$*$-algebra representations with indefinite metric. To this aim
let $\gamma_{K,L}$ be the strongest topology on $\underline{\cal
S}$ s.t. $\forall n\in\N$ the restriction of $\gamma_{K,L}$ to
${\cal S}_n$ is induced by the norms (\ref{2.1eqa}). Let $\gamma$
be te weakest topology on $\underline{\cal S}$ generated by all
$\gamma_{K,L}$. Then we get e.g. by Theorem 3 of \cite{MoSt}:
\begin{theorem}
\label{2.2theo} If $\underline{W}\in\underline{\cal S}'$ fulfills
the condition {\bf (A5')}: $\underline{W}$ is continuous w.r.t.
the topology $\gamma$, then $\underline{W}$ fulfills the HSSC.
\end{theorem}

We note that $\underline{\cal F},\underline{\bar{\cal
F}}:\underline{\cal S}\to\underline{\cal S}$ are
$\gamma$-continuous, thus there is no difference between the
$\gamma$-continuity of $\underline{W}$ and $\underline{\hat W}$.

Topological conditions of this kind obviously are ``linear'' in
the sense that they are preserved under linear combinations. The
only essentially non-linear condition in the set of Axioms
\ref{2.1ax} thus is the cluster property (A6). It is linearized by
an algebraic transformation $\underline{\cal
S}'\ni\underline{W}\mapsto \underline{W}^T\in\underline{\cal S}'$
known as ``truncation''. As we shall see, this transformation
preserves (A2)-(A4), (A7) and transforms (A1) into an equivalent
linear condition. The crucial observation now is that truncation
also preserves the $\gamma$-continuity of $\underline{W}$
\cite{AGW3,Hoff1}. Consequently we can translate the modified
Wightman axioms \ref{2.1ax} into a purely linear set of conditions
for the truncated Wightman functional. For the technicalities we
refer to Appendix A.

\section{Construction of asymptotic states}

In this section we develop a mathematical framework for scattering
in indefinite metric relativistic local QFT. In a certain sense we
go in the opposite direction as the axiomatic scattering theory
with positive metric \cite{Ha,He,Ru} where asymptotic fields are
being constructed first and the scattering amplitudes are
calculated in a second step \cite{He,LSZ}. Here we postulate the
existence of the mixed vacuum expectation values of in- loc- and
out- fields and we then construct these fields using the GNS-like
procedure of Section 2.

Let $\underline{\cal S}^{\rm ext}$ be the ``extended'' Borchers'
algebra over the test function space ${\cal S}_1^{\rm ext}={\cal
S}(\R^d,\C^3)$, which is the space of Schwartz functions with
values in $\C^3$. For $a=$in/loc/out we define $J^a:{\cal
S}_1\to{\cal S}_1^{\rm ext}$ to be the injection of ${\cal S}_1$
into the first/second/third component of ${\cal S}^{\rm ext}_1$,
i.e. $J^{\rm in}f=(f,0,0),J^{\rm loc}f=(0,f,0),J^{\rm
out}f=(0,0,f)$, $f\in{\cal S}_1$. Then $J^a$ uniquely induces a
continuous unital $*$-algebra homomorphism
$\underline{J}^a:\underline{\cal S}\to\underline{\cal S}^{\rm
ext}$ given by $\underline{J}^a=\oplus_{n=0}^\infty J^{a\otimes
n}$.

We also define a suitable ``projection''
$\underline{J}:\underline{\cal S}^{\rm ext}\to\underline{\cal S}$
as the unique continuous unital $*$-algebra homomorphism induced
by  $J:{\cal S}_1^{\rm ext}\to{\cal S}_1$, $J(f^{\rm in},f^{\rm
loc},f^{\rm out})=f^{\rm in}+f^{\rm loc}+f^{\rm out}$.

For simplicity, we only consider the case of only one stable
particle mass $m>0$. Let and $\varphi\in C^\infty_0(\R,\R)$ with
support in $(-\epsilon,\epsilon)$ with $0<\epsilon<m^2$ and
$\varphi(x)=1$ if $-\epsilon/2<x<\epsilon/2$. We define $\chi^\pm
(k)=\theta(\pm k^0)\varphi (k^2-m^2)$ with $\theta$ the Heavyside
step function and we set
\begin{equation}
\label{3.9eqa}
\chi_{t}(a,k)=\left\{
 \begin{array}{ll}
\chi^+(k) e^{-i( k^0-\omega)t}
+\chi^-(k)e^{-i( k^0+\omega)t} & \mbox{ for $a=$in}\\
1 & \mbox{ for $a=$loc}\\
\chi^+(k) e^{i( k^0-\omega)t}
+\chi^-(k)e^{i( k^0+\omega)t} & \mbox{ for $a=$out}
\end{array}
\right.
\end{equation}
We then define $\Omega_t:{\cal S}^{\rm ext}_1\to{\cal S}_1^{\rm ext}$ by
\begin{equation}
\label{3.10eqa}
{\cal F}\Omega_t\bar{\cal F}=\left(\begin{array}{ccc} \chi_t(\mbox{in},k)&0&0\\0&\chi_t(\mbox{loc},k)&0\\0&0&\chi_t(\mbox{out},k)\end{array}\right).
\end{equation}
Next, we introduce the multi parameter $\underline{t}=(t_1,t_2,\ldots),
 t_n=(t_n^1,\ldots,t_n^n), t_n^l\in\R$ and we write $\underline{t}\to+ \infty$ if $t_n^l\to+\infty$
in any order, i.e. first one $t_n^l$ goes to infinity, then the
next etc. . We say that the limit $\underline{t}\to+\infty$ of any
given object exists, if it exists for $t_n^l\to+\infty$ in any
order and it does not depend on the order. We now  define the
finite times wave operator $\underline{\Omega}_{\underline
{t}}:\underline{\cal S}^{\rm ext}\to\underline{S}$ as
\begin{equation}
\label{3.11eqa}
\underline{\Omega}_{\underline{t}}=\underline{J}\circ\oplus_{n=0}^\infty\Omega_{n,t_n}~,~~~\Omega_{0,t_0}=1,~~ \Omega_{n,t_n}=\otimes_{l=1}^n\Omega_{t_n^l}.
\end{equation}
Furthermore, we define the finite times in- and out- wave operators $\underline{\Omega}_{\underline{t}}^{\rm in/out}:\underline{\cal S}\to\underline{\cal S}$ as $\underline{\Omega}_{\underline{t}}\circ \underline{J}^{\rm in/out}$. Up to changes of the time parameter which do not matter in the limit $\underline{t}\to+\infty$, the wave operators $\underline{\Omega}^{(\rm in/out)}_{\underline{t}}$ are $*$-algebra homomorphisms, as can be easily verified from the definitions.
\begin{definition}
\label{3.1def}
{\rm
(i) Let $\underline{W}\in\underline{\cal S}'$ be a Wightman functional s.t. the
functionals $\underline{W}\circ\underline{ \Omega}_{\underline{t}}$ converge in $\underline{\cal S}^{\rm ext '}$
as $\underline{t}\to+\infty$. We then define the form factor functional $\underline{F}\in
 \underline{\cal S}^{\rm ext '}$ associated to $\underline W$ as this limit, i.e.
\begin{equation}
\label{3.14eqa}
\underline{F}=\lim_{\underline{t}\to+\infty}\underline{W}\circ \underline{\Omega}_{\underline{t}}.
\end{equation}
\noindent (ii) The scattering matrix $\underline{S}$ associated to $\underline{W}$ is defined by
\begin{eqnarray}
\label{3.15eqa}
\underline{S}(\underline{f},\underline{g})&=&\underline{F}(\underline{J}^{\rm in}\underline{f}\otimes
\underline{J}^{\rm out}\underline{g})\nonumber \\
&=&\lim_{\underline{t},\underline{t}'\to+\infty}\underline{W}(\underline{\Omega}_{\underline{t}}^{\rm in}\underline{f}\otimes\underline{\Omega}_{\underline{t}'}^{\rm out}\underline{g})~~\forall \underline{f},\underline{g}\in\underline{\cal S}.
\end{eqnarray}
}
\end{definition}

We are now in the position to state a set of conditions which
allow a reasonable definition of the scattering matrix, in- and
out- fields and states in indefinite metric QFT.

\begin{condition}
\label{3.1cond}
{\rm
Let $\underline{W}\in\underline{\cal S}'$. We assume that

\noindent s1) $\underline{W}$ fulfills Axioms  \ref{2.1ax} and
$\underline{W}$ is a theory with a mass gap $m_0>0$, i.e.
$\underline{W}^T(I_{\rm sp}^{m_0})=0$ with $I_{\rm sp}^{m_0}$ the
{\em vector space} generated by $(0,\ldots,0,f_n,0,\ldots)$ with
$\mbox{supp}\hat f_n\subseteq \{ (k_1,\ldots,k_n)\in \R^{dn}:
\exists j,\, 2\leq j\leq n, \mbox{ such that}\sum_{l=j}^nk_l\not
\in \bar V_{m_0}^+ \}$.

\noindent s2) The truncated two point function $W_2^T$ of
$\underline{W}$ is of the form
\begin{equation}
\label{3.6eqa}
\hat W_2^T(k_1,k_2)=\left[ \delta^-_m(k_1)+\int_{m_0}^\infty\delta^-_\mu(k_1)\rho(\mu)d\mu\right] \delta(k_1+k_2)
\end{equation}
with $\rho$ a positive polynomially bounded locally integrable density.

\noindent s3) The form factor functional $\underline{F}$ associated to $\underline{W}$ exists, is Poincar\'e invariant and fulfills the Hilbert space structure condition (HSSC).
}
\end{condition}

The following theorem shows that Condition  \ref{3.1cond} just
implies the LSZ asymptotic condition.

\begin{theorem}
\label{3.1theo}
We suppose that $\underline{W}$ fulfills the Condition \ref{3.1cond}. Then

\noindent (i) There exists a (in general not local) quantum field
theory with indefinite
 metric $(({\cal H},\langle.,.\rangle,\Psi_0),\eta,{\sf U},\Phi)$ over the Borchers algebra
$\underline{\cal S}^{\rm ext}$ such that the statements (i)-(iii) of Theorem \ref{2.1theo} hold.

\noindent (ii) There exist relativistic local quantum fields with
indefinite metric $\phi^{\rm in/loc/out}\linebreak=\Phi\circ
\underline{J}^{\rm in/loc/out}$ over $\underline{\cal S}$  s.t.
$\phi^{\rm in/out}$ are free fields of mass $m$ (for $d\geq4$)and
$\phi=\phi^{\rm loc}$ fulfills the LSZ asymptotic condition,
namely
\begin{equation}
\label{3.16eqa}
\lim_{\underline{t}\to+\infty}\phi(\underline{\Omega}^{\rm in/out}_{\underline{t}}\underline{f})= \phi^{\rm in/out}(\underline{f})
~~\forall \underline{f}\in \underline{\cal S}
\end{equation}
where the limit is taken in ${\sf O}_\eta({\cal D})$.

\noindent (iii) There exist ${\sf U}$-invariant Hilbert spaces ${\cal H}^{\rm in/out}
\subseteq {\cal H}$ defined as $
{\cal H}^{\rm in/out}=\overline{\phi^{\rm in/out}(\underline{\cal S})\Psi_0}$,
s.t. the restriction of $\langle.,.\rangle$ to
${\cal H}^{\rm in/out}$ is positive semidefinite ($d\geq 4$).
\end{theorem}
\noindent {\it Proof.} (i) Except for the spectral property and
Hermiticity, this point of the theorem follows immediately from
s3) and Theorem \ref{2.1theo}. Concerning the spectral property we
note that $\mbox{supp}{\cal F}(W_n\circ \Omega_{n,t_n})\subseteq
\mbox{supp }\hat W_n$. Thus, $\mbox{supp }\hat F_n\subseteq
\mbox{supp }\hat W_n$. Since $\hat W_n$ has the spectral property,
which is actually a restriction on the support of $\hat W_n$, the
spectral property of $\hat F_n$ follows.
$\Phi(I_{\mbox{sp}})\Omega=\{0\}$ now follows from Theorem
\ref{2.1theo}. The Hermiticity follows from the Hermiticity of
$\underline{W}$ the fact that $\underline{\Omega}_{\underline{t}}$
is a $*$-algebra homomorphism (in the sense given above) and that
the limit of Hermitean functionals is Hermitean itself.

(ii) The existence of the fields $\phi^{\rm in/loc/out}$ follows
immediately from point (i) of the theorem, namely from the
existence of the field $\Phi$. That these fields fulfill the
properties of \ref{2.1theo} for $\phi^{\rm in/out}$ follows from
the fact that they are free fields (cf. \cite{SW}) and for
$\phi^{\rm loc}$ this statement by Theorem \ref{2.1theo} follows
from the assumption s1) on $\underline{W}$.

That $\phi^{\rm in/out}$  for $d\geq 4$are free is a consequence of the fact
that the mass gap assumption is fulfilled and thus the truncated
Wightman functionals $W_n^T$ fulfill the strong cluster property
Theorem XI.110 of \cite{RS} Vol. III. Consequently,
$\lim_{\underline{t}\to+\infty}\underline{W}^T(\Omega^{\rm
in/out}_{\underline{t}}\underline{f})=0$ for $f\in\underline{\cal
S}$ with $f_1=0,f_2=0$ follows from Theorem XI.111 in \cite{RS}
Vol. III ( the negative frequency terms which occur in our
framework are just the complex conjugation of some positive
frequency term with the same ``time direction''). The fact that
the locally integrable density $\rho(\mu)d\mu$ does not give a
contribution to the two point function of $\phi^{\rm in/out}$
follows from the Riemann lemma, cf. \cite{RS} Vol. II ( for the
details of the argument, see the proof of Proposition
\ref{4.2prop} below).

In order to prove the ${\sf O}_\eta({\cal D})$-convergence in
Equation (\ref{3.16eqa}), we have to show that $$
\lim_{\underline{t}\to+\infty}\left\langle
\Phi(\underline{f})\Psi_0,\phi(\underline{\Omega}^{\rm
in/out}_{\underline{t}}\underline{g})
\Phi(\underline{h})\Psi_0\right\rangle = \left\langle
\Phi(\underline{f})\Psi_0,\phi^{\rm in/out}(\underline{g})
\Phi(\underline{h})\Psi_0\right\rangle $$ holds for all
$\underline{g}\in\underline{\cal
S},~\underline{f},\underline{h}\in \underline{\cal S}^{\rm ext}$.
Rewriting the left hand side and the right hand side of this
formula in terms of the quantum field $\Phi$ we can verify it
using also s3) by the following calculation
\begin{eqnarray*}
&& \lim_{\underline{t}\to + \infty}\left\langle \Phi(\underline{f})\Psi_0,\Phi(\underline{J}^{\rm loc}\underline{\Omega}^{\rm in/out}_{\underline{t}}\underline{g})
\Phi(\underline{h})\Psi_0\right\rangle\\
 &=&\lim_{\underline{t}\to+\infty}\left\langle \Psi_0,\Phi(\underline{f}^*\otimes\underline{J}^{\rm loc}\underline{\Omega}^{\rm in/out}_{\underline{t}}\underline{g}
\otimes\underline{h})\Psi_0\right\rangle \\
&=&\lim_{\underline{t}\to+\infty}\underline{F}\left(\underline{f}^*\otimes\underline{J}^{\rm
loc}\underline{\Omega}^{\rm in/out}_{\underline{t}}\underline{g}
\otimes\underline{h}\right)\\
&=&\lim_{\underline{t}\to+\infty}\lim_{\underline{s}_1,\underline{s}_2\to+\infty}\underline{W}\left(\underline{\Omega}_{\underline{s}_1}\underline{f}^*\otimes
\underline{\Omega}_{\underline{t}}\underline{J}^{\rm
in/out}\underline{g}\otimes\underline{\Omega}_{\underline{s}_2}\underline{h}\right)\\
&=&\lim_{\underline{t}\to+\infty}\underline{W}\left(\underline{\Omega}_{\underline{t}}(\underline{f}^*\otimes\underline{J}^{\rm
in/out}\underline{g}\otimes\underline{h})\right)\\
&=&\underline{F}\left(\underline{f}^*\otimes \underline{J}^{\rm
in/out}\underline{g}\otimes\underline{h}\right)\\ &=&\left\langle
\Psi_0,\Phi(\underline{f}^*\otimes\underline{J}^{\rm
in/out}\underline{g} \otimes\underline{h})\Psi_0\right\rangle\\
&=& \left\langle\Phi(\underline{f}) \Psi_0,\Phi(\underline{J}^{\rm
in/out}\underline{g}) \Phi(\underline{h})\Psi_0\right\rangle.
\end{eqnarray*}
\noindent (iii) The ${\sf U}$-invariance of ${\cal H}^{\rm
in/out}$ results from the transformation law \linebreak ${\sf
U}(\Lambda,a)\phi^{\rm in/out} (\underline{f}){\sf
U}(\Lambda,a)^{-1}=\phi^{\rm
in/out}(\underline{\alpha}^{-1}_{\{\Lambda,a\}}\underline{f})$ and
the ${\sf U}$-invariance of $\Psi_0$. The transformation law holds
by Theorem \ref{2.1theo} (ii) and s3). That $\langle.,.\rangle$ is
positive semidefinite on ${\cal H}^{\rm in/out}$ follows from the
fact that the dense subspaces  spaces $\phi^{\rm
in/out}(\underline{\cal S})\Psi_0$ are also dense subsets of Fock
spaces over the one particle space ${\cal S}_1$ with positive
semidefinite inner product induced by $W_2^{\rm in/out,\it T}$,
cf. \cite{Bog} p. 288. \kasten

Here we do not give precise conditions for the existence of the
form factor functional, but we refer to the methods of Section 4
and Appendix B where the form factor functional has been
constructed for a special class of models. Looking into the
details of the proof, one notices that what one really requires to
get the existence of this functional is the restriction of
mass-shell singularities to singularities of the type
$\delta_m^\pm(k)$ and $1/(k^2-m^2)$. These are just the
singularities occurring in the Feynman propagator. It therefore
seems to be reasonable that the form factor functional can be
defined in theories where Yang-Feldman equations \cite{YF} hold.
Since for the physicists' common sense Yang-Feldman equations are
an alternative formulation of the LSZ asymptotic condition, to us
it seems that our Condition \ref{3.1cond} does not rule out many
cases of physical interest. The assumption of the existence of the
form factor functional also can not be dropped from Condition
\ref{3.1cond}, since we have to exclude those models from
\cite{AGW6} which have too strong mass shell singularities leading
to divergent Haag-Ruelle like scattering amplitudes.

Finally in this section we translate the Condition  \ref{3.1cond}
into the language of truncated functionals:

\begin{proposition}
\label{3.1prop}
Let $\underline{W}\in\underline{\cal S}'$ be given and $\underline{W}^T$ be the associated truncated Wightman functional.

\noindent (i) $\underline{F}=\lim_{\underline{t}\to +\infty}\underline{W}\circ \underline{\Omega}_{\underline{t}}$ exists if and only if $\underline{ F}^{\tilde T}=\lim_{\underline{t}\to +\infty}\underline{W}^T\circ\underline{\Omega}_{\underline{t}}$ exists. In this case $\underline{F}^{\tilde T}=\underline{F}^T$ (we may thus omit the tilde in the following).

\noindent (ii) $\underline{S}^T(\underline{f},\underline{g})=\underline{F}^T(\underline{J}^{\rm in}\underline{f}\otimes\underline{J}^{\rm out}\underline{g})~\forall \underline{f},\underline{g}\in\underline{\cal S}$.

\noindent (iii) Suppose that $\underline{W}^T$ fulfills s1) (transcribed to the language of truncated Wightman functionals according to Proposition \ref{A.1prop}) and s2) of Condition \ref{3.1cond} and furthermore s3T): $\underline{F}^T\in\underline{\cal S}^{\rm ext'}$ exists, is Poincar\'e invariant and $\gamma$-continuous. Then the associated Wightman functional fulfills s1)-s3) of Condition \ref{3.1cond}.
\end{proposition}

\noindent{\it Proof.} (i) We note that up to the ordering of the
time parameter $\underline{t}$ this statement follows from Lemma
\ref{A.1lem}. But the ordering of $\underline{t}$ does not matter
due to the definition of the limit $\underline{t}\to+\infty$.

(ii) This equation follows by application of Lemma \ref{A.2lem} to $\underline{F}$.

(iii) is a corollary to (i), Proposition \ref{A.1prop} and Theorem \ref{2.2theo}. \kasten

\section{An interpolation theorem}
In this section we  construct a class of quantum fields with
indefinite metric which have a well defined scattering behavior in
the sense of Theorem \ref{3.1theo} and which interpolate a certain
class of scattering matrices. This is being done by a rather
explicit construction of the truncated Wightman functional and the
verification of the conditions given in item (iii) of Proposition
\ref{3.1prop}. The existence of quantum fields with indefinite
metric then follows from Theorem \ref{3.1theo}.

We first recall a well known result of scattering-(S)-matrix
theory following \cite{He,LSZ}: Let us for a moment consider a
quantum field as a operator valued distribution $\phi (x)$ (i.e.
the restriction of the homomorphism $\phi:\underline{\cal
S}\to{\sf O}_\eta({\cal D})$ to ${\cal S}_1$). We assume that
$\phi$ fulfills the LSZ-asymptotic condition $\phi\circ\Omega^{\rm
in/out}_t\to \phi^{\rm in/out}$ as $t\to\infty$ in an appropriate
sense, where the asymptotic fields $\phi^{\rm in/out}$ are  free
fields of mass $m$. Let $\hat \phi^{\rm in/out}(k)$ denote the
Fourier transform of $\phi^{\rm in/out}(x)$. Then the expectation
values of states created by application of the in-fields to the
vacuum $\Psi_0$ with states generated analogously by the
out-fields have the following general shape:
\begin{eqnarray}
\label{4.1eqa} &&\left\langle \hat \phi^{\rm
in}(k_r)\cdots\hat\phi^{\rm in}(k_1)\Psi_0,\hat\phi^{\rm
out}(k_{r+1})\cdots\hat\phi^{\rm out}(k_n)\Psi_0\right\rangle^T
\nonumber\\ &=&  2\pi i~
\underbrace{M_n(-k_r,\ldots,-k_1,k_{r+1},\ldots,k_n)}_{\rm
``transfer~
function''}\underbrace{\prod_{l=1}^n\delta_m^+(k_l)}_{\rm
on-shell~term}~\underbrace{\delta (
\sum_{l=r+1}^nk_l-\sum_{l=1}^rk_l)}_{\rm energy-momentum\atop
conservation~term}\nonumber \\
\end{eqnarray}
where $n\geq 3$, $k_l^0>0,~l=1,\ldots,n$, i.e. all operators
$\hat\phi^{\rm in/out}(k_l)$ are creation operators. Since the in-
and out- fields fulfill canonical commutation relations this is
sufficient to calculate also those expectation values with the
condition on the $k_l^0$ dropped. Here the distribution
$M_n(k_1,\ldots,k_n)\delta(\sum_{l=1}^nk_l)$ is given (up to a
constant) by the Fourier transform of the time-ordered vacuum
expectation values of $\phi(x)$ multiplied by
$\prod_{l=1}^n(k^2_l-m^2)$ and thus is symmetric under permutation
of the arguments and Poincar\'e invariant. By the definition of
the scattering matrix in Section 3, we can equivalently write for
Equation (\ref{4.1eqa})
\begin{eqnarray}
\label{4.2eqa}
&&\hat S_{r,n-r}^T(k_1,\ldots,k_r;k_{r+1},\ldots,k_n)\nonumber\\
&=&  2\pi i~ M_n(k_1,\ldots,k_n)~\prod_{l=1}^r\delta_m^-(k_l)\prod_{l=r+1}^n\delta_m^+(k_l)~\delta ( \sum_{l=1}^nk_l)
\end{eqnarray}
for $n\geq 3$ and $k_l^0<0$ for $l=1,\ldots,r$ and $k_l^0>0$  for
$l=r+1,\ldots,n$. Here we used $\hat\phi^{\rm in/out}(k)=\hat
\phi^{{\rm in/out}[*]}(-k)$.

Given this general form  of the $S$-matrix, one can ask, whether
under some conditions on the transfer functions $M_n$ there exists
an interpolating quantum field $\phi$ s.t. $\phi$ fulfills the LSZ
asymptotic condition and the scattering matrix $\underline{S}$ is
determined by Equation (\ref{4.2eqa}). In the following we give a
(partial) answer to this question for the case of quantum fields
with indefinite metric. First we fix some conditions on the
sequence of transfer functions $M_n$.

\begin{condition}
\label{4.1cond}
{\rm
We assume that $\underline{M}\in\underline{S}'$ fulfills the following conditions:

\noindent I1) $M_n$ is symmetric under permutation  of arguments
and Lorentz invariant (w.r.t. the entire Lorentz group ${\cal
L}$);

\noindent I2) $M_n$ is real, $M_2=1$;

\noindent I3) $M_n$ is a polynomial;

\noindent I4) $\exists L_{\rm max}\in\N_0$ s. t. $\forall n\in\N$
the degree of $M_n(k_1,\ldots, k_n)$ in any of the arguments
$k_1,\ldots,k_n$ is at most $L_{\rm max}$. }
\end{condition}

\begin{remark}
\label{4.1rem} {\rm The ``essentially linear'' set of conditions
given above of course does not imply unitarity of the scattering
matrix, which connects transfer functions of different orders, cf.
\cite{Bog}. Up to now it is not clear, whether in the class of
transfer functions described by Condition \ref{4.1cond} there are
exact solutions to the unitarity condition. ``Approximate''
solutions however are possible due to Proposition \ref{Y.1prop}
below. }
\end{remark}
While the specific properties of the system under  consideration
are encoded in the transfer functions, we also need an input
creating the ``axiomatic structure'', namely the on-shell terms
and the energy-momentum conservation term. In the following we
define a sequence of ``structure functions\footnote{These
functions have nothing to do with the 'structure functions'
describing inelastic scattering in the phenomenology of elementary
particles.}'' with the required properties.

\begin{definition}
\label{4.1def} { \rm For $n\geq 3$ we define the $n$-point
structure function  $G_n$ as the inverse Fourier transform of
$\hat G_n$ given by
\begin{equation}
\label{4.3eqa}
\hat G_n(k_1,\ldots,k_n)=\left\{\sum_{j=1}^n\prod_{l=1}^{j-1}\delta^-_m(k_l){1\over k_j^2-m^2}\prod_{l=j+1}^n\delta^+_m(k_l)\right\}\delta (\sum_{l=1}^nk_l)
\end{equation}
The structure functional $\underline{G}\in\underline{\cal S}'$ is defined by $G_0=0,G_1=0$ and $\hat G_2$ given by Equation (\ref{3.6eqa}).
}
\end{definition}

The structure functions  first have been defined in \cite{AIK}
(for $m=0$), the present form given in Definition \ref{4.1def} was
obtained in \cite{AGW2}. Further properties of the structure
functions are given in \cite{AGW3,AGW4}, see also
\cite{Go2,Go3,Jo}. The following proposition summarizes the
results obtained in these references:

\begin{proposition}
\label{4.1prop} $\underline{G}$ fulfills all properties of a
truncated Wightman functional of a QFT with indefinite metric with
a mass gap $m_0>0$ ( cf. Proposition \ref{A.1prop}, Condition
\ref{3.1cond} s1)).
\end{proposition}

If $\underline{M}$ is a functional which  fulfills Cond.
\ref{4.1cond}, then we define the dot-product of the functionals
$\underline{M}$ and $\underline{\hat G}$ by
$(\underline{M}\cdot\underline{\hat G})_n=M_n\cdot\hat G_n$ where
the multiplication on the right hand side obviously is well
defined, since $\hat G_n$ is a tempered distribution and $M_n$ is
a polynomial.

We now have collected the pieces, which are  being put together in
the following ``interpolation theorem''.
\begin{theorem}
\label{4.1theo}
Let $\underline{G}$ be the structure functional (cf. Definition \ref{4.1def}) 
and let $\underline{M}\in\underline{\cal S}'$ fulfill Condition \ref{4.1cond}. Then

\noindent (i) $\underline{\hat W}^T=\underline{M}\cdot\underline{\hat G}$ 
fulfills the conditions of Proposition \ref{3.1prop} (iii).

\noindent (ii) The truncated S-matrix (cf. Prop. \ref{3.1prop} (ii))
 is determined by Equation (\ref{4.2eqa}).

\noindent (iii) In particular, there exists  a local, relativistic
quantum field $\phi $ with indefinite metric (see Theorem
\ref{2.1theo}) which fulfills the LSZ asymptotic condition
Equation (\ref{3.16eqa}) w.r.t. free fields $\phi^{\rm in/out}$ of
mass $m$ and has scattering behavior de\-ter\-mi\-ned by Equation
(\ref{4.1eqa}). The restriction of the indefinite inner product
$\langle.,.\rangle$ to the Hilbert spaces ${\cal H}^{\rm
in/out}=\overline{\phi^{\rm in/out}(\underline{\cal S})\Psi_0}$ is
positive semidefinite.
\end{theorem}

The rest of this section is devoted to the proof of Theorem
\ref{4.1theo}. Obviously, the item (iii) is a straight forward
application of (i), (ii), Proposition \ref{3.1prop} and Theorem
\ref{3.1theo}\footnote{By a direct calculation as in the proof of (ii) below one
can show that the fields $\phi^{\rm in/out}$ are free fields also for $d=2,3$.
}. Therefore, we only have to check the statements (i)
and (ii). 

\noindent {\bf Proof of statement (i).} {\it Step 1) Verification
of the modified Wightman axioms for $\underline{\hat W}^T$ and
s1),s2)}: (A1T) holds by $G_0=0$ and
$\underline{G}\in\underline{\cal S}'$, cf. Prop. \ref{4.1prop}.
Poincar\'e invariance (A2) follows straightforwardly from the
translation invariance of $\underline{G}$ and Lorentz invariance
of $\underline{G}$ and $\underline{M}$. The (strong) spectral
property (A3) (s1) can be verified by $\mbox{supp} \hat
W_n^T=\mbox{supp} M_n\cdot \hat G_n\subseteq \mbox{supp}\hat
G_n\subseteq \{ (k_1,\ldots,k_n)\in\R^{dn}:\sum_{l=j}^nk_l\in \bar
V_{m_0}^+\mbox{ for }j=2,\ldots,n\}$ for $n\geq 2$ where the last
inclusion holds by Prop. \ref{4.1prop}. Locality (A4) can
equivalently be expressed in terms of the (truncated) Wightman
functions via $\mbox{supp} W_{n,[,]_j}^T\subseteq\{
(x_1,\ldots,x_n)\in\R^{dn}:(x_j-x_{j+1})^2\geq 0\}$ for
$j=1,\ldots,n-1$ where
$W_{n,[,]_j}^T(x_1,\ldots,x_j,x_{j+1},\ldots,x_n)=W^T_n(x_1,\ldots,x_j,x_{j+1},\ldots,x_n)-W^T_n(x_1,\ldots,x_{j+1},x_j,\ldots,x_n)$.
This follows by $$ \mbox{supp } W_{n,[,]_j}^T = \mbox{supp }
M_n(-i{\partial\over\partial x_1},\ldots,-i{\partial\over\partial
x_n})G_{n,[,]_j}\subseteq \mbox{supp }  G_{n,[,]_j} $$ where in
the first step we have made use of the definition of $W_n^T$ and
the symmetry of $M_n$ under permutation of the arguments $j,j+1$,
and in the second step we used that multiplication by a polynomial
in momentum space gives differentiation in position space which is
a local operation. Now the assertion follows from the locality of
$G_n$, cf. Prop. \ref{4.1prop}. The proof of (A5') follows from
the observation that the $\gamma_{c,r}$-continuity of
$\underline{\hat G}$ (which holds for some ${c,r}\in\N$ by Prop.
\ref{4.1prop}) implies the $\gamma_{ c, r+L_{\rm max}}$-continuity
of $\underline{\hat W}^T$ where $L_{\rm max}$ is given in
Condition \ref{4.1cond} I4). (A6T) follows from the strong
spectral property, invariance  and locality, cf. Theorem XI.110 of
\cite{RS} Vol. III. Hermiticity (A7) immediately follows from the
Hermiticity of $\underline{G}$ and the fact that $$
\overline{M_n(-k_n,\ldots,-k_1)}=M_n(k_n,\ldots,k_1)=M_n(k_1,\ldots,k_n)
$$ where we have also used the real valuedness, reflection
invariance and symmetry of $M_n$. But this is just the relation
defining Hermiticity in momentum space. Finally, s2) holds by the
definition of $\underline{G}$ and $M_2=1$.

{\it Step 2) Calculation of the truncated form factor functional
and verification of (A5'), (s3T)}: We proceed as follows: We
define a functional $\underline{F}^G$ and we  prove that this is
the form factor functional associated to $\underline{G}$. To show
this, we require two technical lemmas; their proofs can be found
in Appendix B. The rest of the proof of this step is in a similar
fashion as the preceding paragraph.

 We define the distribution
$\Delta_m \in {\cal S}_1^{\rm ext}$ by the following formula for
the Fourier transform of it's in-, loc- and out- component:
\begin{equation}
\label{4.4eqa}
\hat \Delta_{m} (a,k)=\left\{
\begin{array}{ll}
 -i\pi (\delta_{m}^+(k)-\delta_{m}^-(k)) & \mbox{ for $a=$in}\\
1 \left/(k^2-m^2)\right.                & \mbox{ for $a=$loc}\\
i\pi (\delta_{m}^+(k)-\delta_{m}^-(k)) & \mbox{ for $a=$out}
\end{array}\right.
\end{equation}
Here, as in the  definition of the structure functions, the
singularity $1/(k^2-m^2)$ have to be understood in the sense of
Cauchy's principal value. We now define the functional
$\underline{F}^G$ which turns out to be the form factor functional
associated with $\underline{G}$:
\begin{definition}
\label{4.2def} { \rm The functional
$\underline{F}^G\in\underline{\cal S}^{\rm ext'}$ is defined by
the following formulae for the Fourier transform of the components
$\hat F_n^{G(a_1,\ldots,a_n)}$, $a_l=$in/loc/out, $l=1,\ldots,n$:
$\hat F_0^G=0,~\hat F_1^{G(a_1)}(k_1)=0,$
\begin{equation}
\label{4.5eqa}
\hat F_2^{G(a_1,a_2)}(k_1,k_2)=\left\{ \begin{array}{ll}
\hat G_2(k_1,k_2) & \mbox{for $a_1=a_2=$loc}\\
\delta^-_m(k_1)\delta(k_1+k_2)& \mbox{otherwise}
\end{array} \right.
\end{equation}
and
\begin{equation}
\label{4.6eqa} \hat F_{n}^{G(a_1,\ldots,a_n)}(k_1,\ldots,k_n)
=\left\{\sum_{j=1}^n\prod_{l=1}^{j-1} \delta^-_{m}(k_l) \hat
\Delta_{m}(a_j,k_j)\prod_{l=j+1}^n\delta^+_{m}(k_l)\right\}\delta
(\sum_{l=1}^nk_l).
\end{equation}
}
\end{definition}
That $\underline{F}^G$ is in  $\underline{\cal S}^{\rm ext '}$, as
stated in the Definition \ref{4.2def}, is contained in the
following

\begin{proposition}
\label{4.2prop} $\underline{F}^G$ is the form  factor functional
associated to $\underline{G}$. Furthermore, $\underline{F}^G$ is
Poincar\'e invariant and $\gamma$-continuous.
\end{proposition}

For the proof of this proposition we introduce  the test function
space ${\cal S}_{1,2}=\cap_{L=0}^\infty\overline{\cal
S}_1^{\|\cdot\|_{2,L}}$ (the bar stands for completion) with the
topology of the inductive limit. By ${\cal S}'_{1,2}$ we denote
the topological dual space. It is well-known that $1/(k^2-m^2)$ as
a distribution lies in ${\cal S}_{1,2}'$ (since the Cauchy
principle value in a neighborhood of the singularity is continuous
w.r.t. the $C^1$-norm) and thus $\hat \Delta_m(a,k)\in{\cal
S}_{1,2}'$ for $a=$ in/loc/out. The following two lemmas contain
the analytic part of the proof of Proposition \ref{4.2prop}. For
the proof see Appendix B:

\begin{lemma}
\label{4.1lem} $\lim_{t\to+\infty} {\chi_t(a,k)\over (k^2-m^2)}=\hat \Delta_m(a,k)$ holds in ${\cal S}'_{1,2}$ for $a=${\rm in/loc/out}.
\end{lemma}

\begin{lemma}
\label{4.2lem}
For $f\in{\cal S}_{n}$, $n\geq 3,j=1,\ldots,n,$  let $g_j:\R^d\to\C$ be defined as
\begin{eqnarray}
\label{4.7eqa}
g_j(k_j)&=&\int_{\R^{d(n-1)}}f(k_1,\ldots,k_n)\prod_{l=1}^{j-1}\delta_{m}^-(k_l)\prod_{l=j+1}^n\delta_{m}^+(k_l)\nonumber \\
&\times&\delta(\sum_{l=1}^nk_l)dk_1\cdots dk_{j-1}dk_{j+1}\cdots dk_n.
\end{eqnarray}
Then $g_j\in{\cal S}_{1,2}$ and $\|g_j\|_{2,L}\leq c_L\|f\|_{2,L'}$ for $L\in\N$, $L'=\max\{d,L\}$ and $c_L>0$
sufficiently large.
\end{lemma}

\noindent {\it Proof of Proposition \ref{4.2prop}} We first note
that $\underline{\hat F}^G$ is manifestly Poincar\'e invariant.
The $\gamma$-continuity of $\underline{F}^G$ can be seen as
follows: Let $a_l=$in/loc/out, $l=1,\ldots,n$ be fixed and $f\in
{\cal S}_n$, $n\geq 3$. Then by Lemma \ref{4.2lem} and the fact
that $\hat \Delta_m(a_j,k_j)$ is continuous w.r.t. $\|.\|_{2,L}$
for $L\geq d+1$ (with continuity constant $d_L>0$ sufficiently
large) we get the following estimate:
\begin{eqnarray*}
\left| \hat F_n^{G(a_1,\ldots,a_n)}(f)\right| &=&
\left|\sum_{j=1}^n\int_{\R^d}\hat
\Delta_m(a_j,k_j)g_j(k_j)dk_j\right|\\ &\leq& d_L
\|g_j\|_{2,L}\leq d_Lc_L\|f\|_{2,L}.
\end{eqnarray*}
Thus, if we choose  $L$ sufficiently large s.t. the ``continuous
part'' of $\hat G_2$ (which is determined by $\rho$, cf. Equation
\ref{3.6eqa}) is continuous w.r.t. $\|.\|_{0,L}$, we get that
$\underline{\hat F}^G$ is continuous w.r.t. $\gamma_{2,L}$ and
hence w.r.t. $\gamma$.

To finish the proof we have to show that for $n\in\N_0$
\begin{equation}
\label{4.8eqa}
\lim_{t_n^1,\ldots,t_n^n\to+\infty}\prod_{l=1}^n\chi_{t_n^l}(a_l,k_l)\hat G_n(k_1,\ldots,k_n)=\hat F_n^{G(a_1,\ldots,a_n)}(k_1,\ldots,k_n),
\end{equation}
where $t_n^l\to+\infty$, $l=1,\ldots,n$, in arbitrary order and the limit is being taken in ${\cal S}_n'$. For $n=0,1$ this holds by definition ($G_0,G_1=0$ and $F^G_0,F_G^1=0$). Let $n=2$. For $a_1=a_2=$loc there is nothing to prove since $\chi_t({\rm loc},k)=1$. Let e.g. $a_1=$out and $f\in {\cal S}_2$. Then we get for the left hand side of Equation (\ref{4.8eqa}) smeared out with $f$ for the case first $t_2^1\to+\infty$ and then $t_2^2\to+\infty$
\begin{eqnarray*}
\ldots &=&\lim_{t_2^2\to+\infty}\lim_{t_2^1\to+\infty}
\int_{\R^d}\left[ \delta_m^-(k)+\int_{m_0}^\infty
\delta^-_\mu(k)\rho(\mu)d\mu\right]\\ &\times&
e^{i(k^0+\omega)t_2^1} \chi^-(k) \chi_{t_2^2}(a_2,-k)f(k,-k)~dk\\
&=& \int_{\R^d}\delta^-_m(k)f(k,-k)~dk
+\lim_{t_2^2\to\infty}\lim_{t_2^1\to+\infty} \int_{m_0}^\infty
e^{i(\omega-\omega_\mu) t_2^1}\\ &\times&
\left[\int_{\R^{d-1}}f((-\mu,{\bf k}),(\mu,-{\bf
k}))\varphi(\mu^2-m^2)\chi_{t_2^2}(a_2,(-\mu,{\bf k})){d{\bf
k}\over 2\omega_\mu }\right] \rho(\mu)d\mu
\end{eqnarray*}
Here $\omega_\mu=\sqrt{|{\bf k}|^2+\mu^2}$. We want to show that
the limit of the second integral vanishes. To do this, we note
that the expression in the brackets $[\ldots]$ defines a smooth
and fast falling (for $\mu\to+\infty$) function in $\mu$ and the
change of variables $\mu\to \xi=\omega-\omega_\mu$ is smooth (with
polynomially bounded determinant) since $m_0>0$. Thus, the second
integral can be written as the Fourier transform evaluated at
$t_2^1$ of a $L^1(\R)$-function in the variable $\xi$ (which might
depend on $t_2^2$). By the lemma of Riemann-Lebesgue (cf. Theorem
IX.7 \cite{RS} Vol. II ), the Fourier transform of such a function
vanishes at infinity. Thus, the second integral vanishes. If we
first take the limit $t_2^2\to+\infty$ and then $t_2^1\to\infty$,
we can distinguish two cases: If $a_2=$loc, the second integral
does not depend on $t_2^2$ and we can thus take the limit
$t_2^1\to+\infty$ as before. If $a_2\not=$loc we get by an
argument which is analogous to the one given above, that the limit
$t_2^2\to+\infty$ of the second integral on the r.h.s. vanishes.
This proves Equation (\ref{4.8eqa}) for the case $n=2$.

Let thus $n\geq 3$. Using the fact that $\chi_t(a,k)\delta^\pm_m(k)=\delta^\pm_m(k)$ we get for the left hand side of (\ref{4.8eqa}) smeared out with $f\in{\cal S}_n$:
$$
\ldots=\sum_{j=1}^n\lim_{t_n^j\to+\infty}\int_{\R^d}{\chi_{t_n^j}(a_j,k_j)\over k_j^2-m^2}g_j(k_j)~dk_j.
$$
where we have used the notation introduced in Lemma \ref{4.2lem}. Using now that by Lemma \ref{4.2lem} $g_j\in{\cal S}_{1,2}$ we get by Lemma \ref{4.1lem} for the right hand side of this equation
$$
 \ldots=\sum_{j=1}^n\int_{\R^d}\hat \Delta_m (a_j,k_j)g_j(k_j)~dk_j.
$$
But this is just the right hand side of Equation (\ref{4.8eqa}) smeared out with $f$. \kasten

Similar as above, we define the dot-product $\underline{M}\cdot\underline{\hat F}^G\in\underline{\cal S}^{\rm ext'}$ of $\underline{\hat F}^G$ with $\underline{M}$ via $(\underline{M}\cdot\underline{\hat F}^G)_n^{(a_1,\ldots,a_n)}=M_n\cdot \hat F_n^{G(a_1,\ldots,a_n)}$. We then get from Proposition \ref{4.2prop} by a simple use of duality and the same arguments as in step 1):

\begin{corrollary}
\label{4.1cor}
$\underline{F}^T=\underline{\bar{\cal F}}(\underline{M}\cdot\underline{\hat F}^G)$ exists, is Poincar\'e invariant and $\gamma$-con\-tinu\-ous.
\end{corrollary}

\noindent {\bf Proof of statement (ii)} Let $n\geq 3$, $1\leq
r\leq n-1$, $k_1^0,\ldots,k_r^0<0$ and $k_{r+1}^0,\ldots,k_n^0>0$.
Then by Corollary \ref{4.1cor}
\begin{eqnarray*}
&&\hat S_{r,n-r}^T(k_1,\ldots,k_r;k_{r+1},\ldots,k_n) = \hat F_n^{T({\rm in,\ldots,in, out,\ldots,out})}(k_1,\ldots,k_n)\\
&=& M_n(k_1,\ldots,k_n)\hat F_n^{G ({\rm in,\ldots,in,out,\ldots,out})}(k_1,\ldots,k_n),
\end{eqnarray*}
where the ``in'' is being repeated $r$ times and the ``out'' $n-r$ times. Inserting (\ref{4.6eqa}) into this expression we get
\begin{eqnarray*}
&&M_n(k_1,\ldots,k_n)\left\{ \sum_{j=1}^r\prod_{l=1}^{j-1}\delta^-_{m}(k_l) \hat \Delta_{m}({\rm in},k_j)\prod_{l=j+1}^n\delta^+_{m}(k_l)\right.\\
&+&\left. \sum_{j=r+1}^n\prod_{l=1}^{j-1}\delta^-_{m}(k_l) \hat \Delta_{m}({\rm out},k_j)\prod_{l=j+1}^n\delta^+_{m}(k_l)\right\} \delta (\sum_{l=1}^n k_l).
\end{eqnarray*}

Using  the assumption $k_1^0,\ldots, k_r^0<0$ and $k_{r+1}^0,\ldots,k_n^0>0$ for $j=r+1,\ldots,n$
we see that in the first sum only the term $j=r$ gives a non vanishing contribution whereas in the second sum all terms vanish except for the term $j=r+1$. Inserting the expression (\ref{4.4eqa}) and using $k^0_r<0$ and $k_{r+1}^0>0$ we see that the $\delta^+_m(k_r)$-term in $\hat \Delta_m ({\rm in},k_r)$ gives no contribution and this is also true for the $\delta^-_m(k_{r+1})$-term in $\hat \Delta_m({\rm out},k_{r+1})$. We thus get for the above expression
\begin{eqnarray*}
&&M_n(k_1,\ldots,k_n)\left\{ \prod_{l=1}^{r-1}\delta^-_{m}(k_l) [i\pi ~ \delta^-_m(k_r)]\prod_{l=r+1}^n\delta^+_{m}(k_l)\right.\\
&+&\left. \prod_{l=1}^{r}\delta^-_{m}(k_l) [i\pi~ \delta^+_m(k_{r+1})]\prod_{l=r+2}^n\delta^+_{m}(k_l)\right\} \delta (\sum_{l=1}^n k_l)\\
&=&  2\pi i~ M_n(k_1,\ldots,k_n)~\prod_{l=1}^r\delta_m^-(k_l)\prod_{l=r+1}^n\delta_m^+(k_l)~\delta ( \sum_{l=1}^nk_l).
\end{eqnarray*}
This finishes the proof of Theorem \ref{4.1theo}.
\section{Approximation of arbitrary scattering amplitudes}

Here we want to discuss the approximation of a given
(``reference'') set of transfer functions (cf. equation
(\ref{4.1eqa})) $\underline{R}$ with polynomial transfer functions
$\underline{M}$. For $\underline{R}$ we assume full Lorentz
invariance (including reflections) and symmetry under permutation
of the arguments, which is motivated from the LSZ formalism(see
Section 4). Furthermore, we assume that the $R_n$ are continuous,
real functions\footnote{In general scattering amplitudes are
analytic functions on a ``cut'' neighborhood of the on-shell
region and therefore can have discontinuities or singularities on
these ``cuts'', cf. \cite{E,We}. Therefore, we do not consider
$\underline{R}$ as the transfer functions of some ``real'' theory,
but as a set of ``measurement data''. Then, the requirement of
realness can be justified by the fact that only the square modulus
of $R_n$ enters in the measurable transition probabilities and
continuity can be understood in the sense that $R_n$ was obtained
by some continuous interpolation of a discrete set of
measurements.}.  Since the models of Section 4 have polynomial
transfer functions, which grow very fast for large energy
arguments and therefore have a somehow 'bad' high energy
behaviour, we only consider scattering experiments with maximal
energy $E_{\rm max}>0$, which can be chosen arbitrarily large.

By $Q_n({E_{\rm max}})$, we denote the set of points in energy-momentum space which can be reached by a scattering experiment of maximal energy $E_{\rm max}$:
\begin{eqnarray}
\label{Y.1eqa}
&&\bigcup_{1\leq r\leq n-1}\Big\{ (k_1,\ldots,k_n)\in\R^{dn}: k_l^2= m^2, l=1,\ldots,n; k_1^0,\ldots,k_r^0<0,\nonumber \\
&&k_{r+1}^0,\ldots,k_n^0> 0, \sum_{l=r+1}^nk_l^0\leq E_{\rm max},\sum_{l=1}^nk_l=0\Big\}.
\end{eqnarray}
It is easy to verify that for $E_{\rm max}<\infty$, $Q_{n}(E_{\rm max})$ is compact and that $Q_n(E_{\rm max})\linebreak =\emptyset$ for $n>E_{\rm max}/m$. We say that $\underline{M}$ approximates $\underline{R}$ for energies smaller than $E_{\rm max}$ up to an error $\epsilon>0$, if for $n\in\N$ $|M_n(k_1,\ldots,k_n)-R_n(k_1,\ldots,k_n)|<\epsilon$ holds $\forall (k_1,\ldots,k_n)\in Q_{n}(E_{\rm max})$. We then get

\begin{proposition}
\label{Y.1prop} Let $\underline{R}$ be a real, fully Lorentz
invariant and symmetric functional consisting of continuous
functions. For any error parameter $\epsilon>0$ arbitrarily small
and any energy cut-off parameter $E_{\rm max}>0$, there exists a
functional $\underline{M}$ which fulfills the Conditions
\ref{4.1cond} and which approximates $\underline{R}$ for energies
smaller than $E_{\rm max}$ up to an error $\epsilon$ (in the sense
given above).

In particular, there exists a QFT with indefinite metric in the
class of QFTs given in Theorem \ref{4.1theo} with scattering
behavior which for energies smaller than $E_{\rm max}$ differs
from the data $\underline{R}$ at most by an error $\epsilon$.
\end{proposition}

We start the proof with a technical lemma:

\begin{lemma}
\label{Y.1lem} Let $R_n:(\bar V_{m_0}^+\cup \bar
V_{m_0}^-)\times\R^{d(n-1)} \to \R$  be continuous and invariant
under the full Lorentz group ${\cal L}$. Then there exists a
continuous function $V_n:\R^{n(n+1)/2}\to\R$ s.t. $$
R_n(k_1,\ldots,k_n)=V_n(k_1^2,k_1\cdot k_2,k_2^2,\ldots,k_1\cdot
k_n,k_2\cdot k_n,\ldots, k_n^2). $$
\end{lemma}

\noindent {\it Sketch of the Proof.} Let $\pi:(\bar V_{m_0}^+\cup
\bar V_{m_0}^-)\times\R^{d(n-1)}/{\cal L} \to \R^{n(n+1)/2}$ be
defined by ${\cal L}\bar k={\cal L}(k_1,\ldots,k_n)\to
(k_1^2,k_1\cdot k_2,k_2^2,\ldots,k_1\cdot k_n,k_2\cdot k_n,\ldots,
k_n^2)=(q_{1,1},\ldots,q_{n,n})=\bar q$. We want to define $V_n$
on the image of $\pi$ as $R_n\circ \pi^{-1}$. Hence we have to
show that $\bar{k},\bar{k}'\in \pi^{-1}(\bar{q})$ are in the same
orbit of ${\cal L}$ in $\R^{dn}$.

First, we can apply a Lorentz boost (possibly in connection with
time reflection) which maps $k_1$ ($k_1'$) to $(\sqrt{k_1^2},{\bf
0})$. Then, in this new frame of reference the zero components of
$k_l, l=2,\ldots,n$ are given by $k_l\cdot k_1/\sqrt{k_1^2}$ (for
$k_l'$ we proceed analogously). Since the zero components are
known, also scalar products of the ${\bf k}_l$ (${\bf k}'_l$) are
known in this new frame which fixes distances of 'points' from the
origin and 'angles' of the 'rigid body' spanned by the ${\bf k}_l$
(${\bf k}_l'$) in $\R^{d-1}$. But then there is a orthogonal
transformation on $\R^{d-1}$ moving the 'rigid body' spanned by
${\bf k}_l$ onto the one spanned by the ${\bf k}_l'$. Hence
$\bar{k}$ and $\bar{k}'$ are in the same orbit.

Furthermore, the mapping $V_n$ is continuous on the set $\mbox{Ran
}\pi$. This follows from the fact that one can construct a
reference vector $\bar{r}(\bar{q})\in\R^{dn}$ corresponding to
fixing the zero component and a 'standard orientation' for the
'rigid body' which depends smoothly of $\bar{q}$. Thus, for
$\bar{q}_n\to\bar{q}'$ in $\mbox{Ran }\pi$ we get
$\bar{r}_n\to\bar{r}'$ and thus $V_n(\bar{q}_n)=R_n(\bar{r}_n)\to
R_n(\bar{r}')=V_n(\bar{q}')$. Since $\mbox{Ran }\pi$ is closed in
$\R^{n(n+1)/2}$, there exists a continuous extension of $V_n$ to
$\R^{n(n+1)/2}$. \kasten

\noindent {\it Proof of Proposition \ref{Y.1prop}} We use  the
same notations as in the proof of Lemma \ref{Y.1lem}. Note that
$\pi(Q_n(E_{\rm max}))$ is compact since $\pi$ is continuous and
$Q_n(E_{\rm max})$ is compact. Thus, for $\epsilon>0$ by the
Stone-Weierstrass theorem there exists a polynomial $p_n$ such
that $|p_n(\bar{q})-V_n(\bar{q})|<\epsilon$ $\forall \bar{q}\in
\pi (Q_n(E_{\rm max}))$. Let thus
$M_n(\underline{k})=p_n(\pi(\bar{k}))$, then
$|M_n(\bar{k})-R_n(\bar{k})|<\epsilon$ $\forall \bar{k}\in
Q_n(E_{\rm max})$. Furthermore, there is no problem to assume that
$M_n$ is real and symmetric under exchange of variables, since if
this is not the case we can replace $M_n$ with $\mbox{Re}M_n$ and
symmetrize without changing the approximation properties.

By construction $M_n$ is invariant under the full Lorentz group.
It remains to show that the uniform bound in the degree of
$M_n(k_1,\ldots,k_n)$ can be obtained. But this follows from
$Q_n(E_{\rm max})=\emptyset$ for $n>E_{\rm max}/m$, which means
that we can chose $\{M_n\}_{n>E_{\rm max}/n}$ as arbitrary real,
symmetric and Lorentz invariant polynomials with uniform bound.
\kasten

By Proposition \ref{Y.1prop},  there is no 'falsification'  based
on scattering experiments for the statement that the ``true''
theory explaining a set of measurements $\underline{R}$ is in the
class of models given in Theorem \ref{4.1theo} (note that
$\langle.,.\rangle$ is positive semidefinite on the asymptotic
states, thus there is no problem with the probability
interpretation of such experiments). Of course, we do not consider
this as a serious physical statement. Instead, we think that this
result emphasizes the importance of structural aspects (as e.g. a
``good'' high energy behavior, ``exact'' unitarity), which might
go beyond an explicit and exact measurability.

\appendix
\section{Truncation of (bi-) linear functionals on Bor\-chers' algebra}
We introduce the following notation: Let $\lambda_l=(\lambda^1_l,\ldots,\lambda^j_l)\subseteq (1,\ldots,n)$ where the inclusion means that $\lambda_l$ is a subset of $\{1,\ldots,n\}$ and the natural order of $(1,\ldots,n)$ is preserved. Let ${\cal P}(1,\ldots,n)$ denote the collection of all partitions of $(1,\ldots,n)$ into disjoint sets $\lambda_l$, i.e. for $\lambda\in {\cal P}(1,\ldots,n)$ we have $\lambda=\{\lambda_1,\ldots,\lambda_r\}$ for some $r$ where $\lambda_l\subseteq (1,\ldots,n)$, $\lambda_l\cap\lambda_{l'}=\emptyset$ for $l\not = l'$ and $\cup_{l=1}^r\lambda_l=\{ 1,\ldots,n\}$. Given a Wightman functional $\underline{W}\in\underline{\cal S}'$ and $\lambda_l=(\lambda_l^1,\ldots,\lambda_l^j)$, we set $W(\lambda_l)=W_j(x_{\lambda_l^1},\ldots,x_{\lambda_l^j})$.

With this definition at hand we can recursively define the truncated Wightman functional $\underline{W}^T\in\underline{\cal S}'$ associated to $\underline{W}\in\underline{\cal S}'$ via $W_0^T=0$ and
\begin{equation}
\label{A.1eqa}
W(1,\ldots,n)=\sum_{\lambda\in{\cal P}(1,\ldots,n)}\prod_{l=1}^{|\lambda|} W^T(\lambda_l)~,n\in\N,
\end{equation}
where $|\lambda|$ is the number of sets $\lambda_l$ in $\lambda$. We have the following proposition on the properties of $\underline{W}^T$:

\begin{proposition}
\label{A.1prop} $\underline{W}$ fulfills the axioms  \ref{2.1ax}
(A1)-(A4),(A5'),(A6) and (A7) if and only if $ \underline{W}^T$
fulfills (A1T): $W_0=0,\underline{W}^T\in\underline{\cal S}'$,
(A2)-(A4), (A5'),(A7) and (A6T): $\lim_{t\to\infty}\underline{W}^T
(\underline{f}\otimes\underline{\alpha}_{\{1,ta\}}\underline{g})=0$
for $a\in\R^d$ space like and
$\underline{f},\underline{g}\in\underline{S}$ with $f_0=g_0=0$.
\end{proposition}
\noindent{\it Proof.} The equivalence of (A1)/(A2)-(A4)/(A7) for
$\underline{W}$ $\Leftrightarrow$ (A1T)/(A2)-(A4)/\linebreak (A7)
for $\underline{W}^T$ can be found e.g. in \cite{Bog} pp. 492-493.
(A6) for $\underline{W}$ $\Leftrightarrow$ (A6T) for
$\underline{W}^T$ is well-known, for a detailed proof cf.
\cite{AGW2} section 4. (A5') for $\underline{W}$ $\Leftrightarrow$
(A5') for $\underline{W}^T$ is proven in \cite{AGW3,Hoff1}.
\kasten

For continuous operators $A:{\cal S}_1\to{\cal S}_1$ we define
$A_n=A^{\otimes n}$, $A_0=1$ and we set
$\underline{A}^\otimes:\underline{\cal S}\to\underline{\cal S}$
setting $\underline{A}^\otimes=\oplus_{n=0}^\infty A_n$. We get

\begin{lemma}
\label{A.1lem} Let $A:{\cal S}_1\to{\cal S}_1$ be linear and
continuous. Then $\underline{W}^T\circ
\underline{A}^\otimes=(\underline{W}\circ\underline{A}^\otimes)^T~\forall
\underline{W}\in \underline{\cal S}'$.
\end{lemma}

Since the scattering matrix can be considered as a bilinear
functional on the Borchers' algebra, we require a definition of
truncation for these objects. By the Schwartz kernel theorem it is
clear that there is a one to one correspondence of the bilinear
functionals $\underline{S}$ on $\underline{\cal S}$ with sets of
tempered distributions $\{S_{n,m}\}_{n,m\in\N_0}$ where
$S_{n,m}\in{\cal S}_{n+m}$ and
$\underline{S}(\underline{f},\underline{g})=\sum_{n,m=0}^\infty
S_{n,m}(f_n\otimes g_m)$. For
$\lambda_l=(\lambda_l^1,\ldots,\lambda_l^r) \subseteq
(1,\ldots,n),\nu_j=(\nu_j^1,\ldots,\nu_j^q)\subseteq
(n+1,\ldots,n+m)$ we define
$S(\lambda_l,\nu_j)=S_{r,q}(x_{\lambda^1_l},
\ldots,x_{\lambda^r_l};x_{\nu^1_j},\ldots, x_{\nu^q_j})$. With
this notation we define recursively the truncated bilinear
functional $\underline{S}^T$ associated with $\underline{S}$ via
\begin{equation}
\label{A.2eqa}
S(1,\ldots,n;n+1,\ldots,n+m)=\sum_{\lambda\in{\cal P}(1,\ldots,n+m)}\prod _{l=1}^{|\lambda|}S^T(\lambda_l^<,\lambda_l^>).
\end{equation}
Here $\lambda_l^<=\lambda_l\cap(1,\ldots,n)$ and $\lambda_l^>=\lambda_l\cap(n+1,\ldots,n+m)$.

The truncation of linear and bilinear functionals  is related as
follows: Let $\imath_\otimes$ be the injection of linear
functionals into the bilinear functionals on $\underline{\cal S}$
given by $\imath_\otimes \underline
{W}(\underline{f},\underline{g})=\underline{W}(\underline{f}\otimes\underline{g})~\forall
\underline{f},\underline{g}\in\underline{\cal S}$. Then we get
from these definitions:

\begin{lemma}
\label{A.2lem} $\imath_\otimes\underline{W}^T=(\imath_\otimes\underline{W})^T~\forall \underline{W}\in\underline{\cal S}'$.
\end{lemma}

\section{Proof of Lemma 4.8 and Lemma 4.9}

\noindent {\it Proof of Lemma 4.8} We begin the proof of Lemma
\ref{4.1lem} with two auxiliary lemmas (for the definition of
${\cal S}_{1,2},{\cal S}_{1,2}^{'}$ cf. Sect. 4).

\begin{lemma}
\label{B.1lem}
The Fourier transform is a continuous mapping from $L^{1'}(\R,\C)$ to ${\cal S}_{1,2}'(\R,\C)$.
\end{lemma}
\noindent {\it Proof.} We prove that ${\cal F}:{\cal
S}_{1,2}(\R,\C)\to L^1(\R,\C)$ is continuous. Then the statement
of the lemma follows by duality. The stated continuity property is
established by the following estimate:
\begin{eqnarray}
\label{B.1eqa}
\|{\cal F}f\|_{L^1(\R,\C)}&=&(2\pi)^{-1/2}\int_{\R}\left|\int_{\R}e^{-i\xi t}f(\xi)~d\xi\right| dt\nonumber \\
&=&(2\pi)^{-1/2}\int_{\R}\left|\int_{\R}e^{-i\xi t}\left(1-{d^2\over d\xi^2}\right)f(\xi)~d\xi\right| {dt\over 1+t^2}\nonumber \\
&\leq&\pi (2\pi)^{-1/2}\int_{\R}\left|\left(1-{d^2\over d\xi^2}\right)f(\xi)\right|~d\xi \leq c \|f\|_{2,2},
\end{eqnarray}
for a sufficiently large constant $c>0$ (here we have used $\int_{\R}dt/(1+t^2)=\pi$).\kasten

Let $1/\xi$ be defined as the Cauchy principal value of the
function $1/\xi$ and the distribution $1/(\xi\pm i0)$ as the
boundary value of $1/(\xi\pm i\epsilon)$ for $\epsilon \to +0$.
$1/\xi$ and $1/(\xi\pm i 0)$ are related via the
Sokhotsky-Plemelji formula
\begin{equation}
\label{B.2eqa}
{1\over \xi\pm i0}={1\over \xi}\mp i\pi\delta(\xi)~,
\end{equation}
cf. \cite{Co} p. 45.  These distributions can be understood as
elements on ${\cal S}_{1,2}'(\R,\C)$, since the Cauchy principle
value is defined on ${\cal S}_{1,2}(\R,\C)$ by \cite{Co} p. 44 and
the delta distribution of course also is defined on this space.
Furthermore, the Fourier transform (in ${\cal S}_1'(\R,\C)$) of
the step function $1_{\{ 0\leq \pm s\}}$ is
\begin{equation}
\label{B.3eqa}
{\cal F}_s(1_{\{ 0\leq\pm s\}}(s))(\xi)=(2\pi)^{-1/2}{\mp i\over \xi\mp i0}~,
\end{equation}
see \cite{Co} p. 94.

\begin{lemma}
\label{B.2lem}
$\lim_{t\to+\infty}e^{\pm i\xi t}/\xi=\pm i\pi \delta (\xi)$
in ${\cal S}_{1,2}'(\R,\C)$.
\end{lemma}
\noindent {\it Proof.} We note that
\begin{eqnarray*}
\lim_{t\to+\infty}{1\over \xi}e^{\pm i\xi t}&=&{1\over \xi}\lim_{t\to +\infty}\left[ \int_0^t{d\over ds}e^{\pm i\xi s} ~ ds+1\right]\\
&=& \pm i (2\pi)^{1/2}\lim_{t\to+\infty}\bar{\cal F}_s(1_{\{0\leq s\leq t\}})(\pm\xi)+{1\over \xi}.
\end{eqnarray*}
Since by Lemma \ref{B.1lem} the (inverse) Fourier transform $\bar {\cal F}_s$ is continuous from $L^{1'}(\R,\C)$ to ${\cal S}_{1,2}'(\R,\C)$ and
$1_{\{0\leq s\leq t\}}(s)\to 1_{\{0\leq s\leq\infty\}}(s)$ as $t\to +\infty$ in \linebreak $L^{1'}(\R,\C)$, we get for the r.h.s.  of the above equation
using also the formulae (\ref{B.2eqa}), (\ref{B.3eqa}):
\begin{eqnarray*}
\cdots &=& \pm i (2\pi)^{1/2}\bar {\cal F}_s(1_{\{0\leq s\}})(\pm \xi)+{1\over \xi}\\
&=& \mp \left[\pm {1\over \xi}-i\pi \delta(\xi)\right]+{1\over \xi} =\pm i\pi\delta(\xi).
\end{eqnarray*}
\kasten

Now we are in the position to prove Lemma \ref{4.1lem}. We only prove the lemma for $a=$out. The case $a=$in is in the same manner and the case $a=$loc is trivial.

We note that the function $f$ in the expression $\chi_t({\rm out},k)f(k)$, $f\in {\cal S}_{1,2}$, can be written as a sum of a function $f_1$ with $\mbox{supp }f_1\subseteq \R^d_+=(0,\infty)\times \R^{d-1}$ and $\mbox{supp }f_2\subseteq \R^d_-=(-\infty,0)\times\R^{d-1}$. Here we only deal with the ``positive frequency part'' $f_1$, and identify $f_1$ with the expression $\chi^+f_1$, which does not change $f_1$ on the mass shell. Furthermore, we omit the index $1$ in the following. Let thus $f\in {\cal S}_{1,2}$ with $\mbox{supp }f\subseteq \R^d_+$. Then
\begin{eqnarray*}
\lim_{t\to+\infty}\int_{\R^d}{e^{i(k^0-\omega)t}\over
k^2-m^2}f(k)~dk &=&
\lim_{t\to+\infty}\int_{\R^{d-1}}\left[\int_{\R}{e^{
i(k^0-\omega)t}\over k^2-m^2}f(k)~dk^0\right] ~d{\bf k}\\
&=&\lim_{t\to+\infty}\int_{\R^{d-1}}\left[\int_{\R}{e^{ i\xi
t}\over \xi }{f(\xi+\omega,{\bf k})\over \xi+2\omega}~d\xi\right]
~d{\bf k}
\end{eqnarray*}
where we have used the change of variables $k^0\to \xi=k^0-\omega$
in the last step. We note that $f(\xi+\omega,{\bf k})/(
\xi+2\omega)$ is in ${\cal S}_{1,2}(\R,\C)$ for ${\bf k}\in
\R^{d-1}$ since the denominator $(\xi +2\omega)$ is smooth on the
support of $f(\xi+\omega,{\bf k})$. Thus, if we can interchange
the $\int_{\R^{d-1}}\cdots d{\bf k}$ integral and the limit
$\lim_{t\to+\infty}$ we get the formula of Lemma \ref{4.1lem} by
$\delta_{m}^+=\delta(k^0-\omega)/2\omega$ and application of Lemma
\ref{B.2lem}.

Let $h\in{\cal S}_{1,2}(\R^d,\C)$. We define $g_t({\bf
k})=\int_{\R}{e^{i\xi t}\over \xi}h(\xi,{\bf k})~d\xi$. Using the
product formula for the inverse Fourier transform on ${\cal
S}'(\R,\C)$ we get
\begin{eqnarray*}
|g_t({\bf k})|&=&2\pi \left| \left[\bar {\cal F}_\xi({1\over
\xi})\ast \bar{\cal F}_\xi(h(\xi,{\bf k}))\right](t)\right|\\
&=&2\pi \left|\int_{\R} i(\pi-1_{\{t-x>0\}}(t-x))\bar {\cal
F}_\xi(h(\xi,{\bf k}))(x)~dx\right|\\ &\leq&2\pi (\pi+1) \int_{\R}
\left|\bar {\cal F}_\xi (h(\xi,{\bf k}))(x)\right| dx\\ &\leq&
c_1\sup_{\xi \in \R,0\leq l\leq 2} |(1+\xi^2){d^l\over
d\xi^l}h(\xi,{\bf k})|\leq c_2{\| h\|_{2,d}\over (1+|{\bf
k}|^2)^{d/2}},
\end{eqnarray*}
for  some $c_1,c_2>0$ sufficiently large. Here we made use of the
estimate (\ref{B.1eqa}) and we also applied the formulae
(\ref{B.2eqa}) and (\ref{B.3eqa}). But this estimate shows that
there is an integrable majorant for $g_t, t\in\R,$, namely
$c/(1+|{\bf k}|^2)^{d/2}$, and we may therefore interchange the
limit $t\to+\infty$ and the integral over $\R^{d-1}$ by the
theorem of dominated convergence.

\

\noindent {\it Proof of Lemma 4.9} For notational convenience we
only prove the lemma for $j=1$. The proof for $j=2,\ldots,n-1$ can
be carried out analogously. By integrating over the variables
$k_2^0,\ldots,k_n^0$ and over ${\bf k}_2$ we obtain for the right
hand side of (\ref{4.7eqa})
\begin{eqnarray*}
&&\int_{\R^{(d-1)(n-2)}}{f(k_1,(\omega_{2},-{\bf
k}_1-\sum_{l=3}^n{\bf k}_l),(\omega_{3},{\bf
k}_3),\ldots,(\omega_{n},{\bf k}_n)) \over
\prod_{l=2}^n\omega_{l}}\\
&\times&\delta(k_1^0+\sum_{l=2}^n\omega_{l})~d{\bf k}_3\cdots
d{\bf k}_n.
\end{eqnarray*}
Here $\omega_{2}=(|{\bf k}_1+\sum_{l=3}^n{\bf k}_l|^2+m^2)^{1/2}$.
We set $$ h(k_1,{\bf k}_3,\ldots,{\bf k}_n)=
{f(k_1,(\omega_{2},-{\bf k}_1-\sum_{l=3}^n{\bf
k}_l),(\omega_{3},{\bf k}_3),\ldots,(\omega_{n},{\bf
k}_n))\over\prod_{l=2}^n2\omega_{l}}. $$ and we get that
$h(k_1,{\bf k}_3,\ldots,{\bf k}_n)\in{\cal
S}_{1,2}(\R^{d+(d-1)(n-2)},\C)$ and $\|h\|_{2,L}\leq \linebreak
c_L \|f\|_{2,L}$ for some $c_L>0$.
 We thus have to show, that for such $h$
\begin{equation}
\label{B.4eqa} g(k)=\int_{\R^{(d-1)(n-2)}}h(k,{\bf
k}_3,\ldots,{\bf k}_n)\delta(\rho({\bf k},{\bf k}_3,\ldots,{\bf
k}_n)+k^0)~d{\bf k}_3\cdots d{\bf k}_n
\end{equation}
defines a ${\cal S}_{1,2}$-function  $g$ and that $\|
g\|_{2,L}\leq c_L\|h\|_{2,L'}$ for $c_L>0$ suf\-fi\-ci\-ent\-ly
large, where we have set $\rho({\bf k},{\bf k}_3,\ldots,{\bf
k}_n)=\sum_{l=2}^n\omega_{l}$.

Using a smooth partition of unity which has bounded derivatives we
can write $h$ as a sum of functions $h_1,h_2,h_3$ where on the
support of $h_1$ we have $|{\bf k}|>1$ and $|\sum_{l=3}^n{\bf
k}_l|>1$, on the support of $h_2$ we have $|{\bf k}|<2,
|\sum_{l=3}^n{\bf k}_l|>1$ and on the support of $h_3$ we have
$|{\bf k}|<2, |\sum_{l=3}^n{\bf k}_l|<2$. By the boundedness of
derivatives of the partition of unity, $\|h_j\|_{2,L}\leq c_L
\|f\|_{2,L}$ holds for $L\in\N_0,j=1,2,3$ and sufficiently large
$c_L>0$. We denote the functions associated to $h_j$ via Equation
(\ref{B.4eqa}) by $g_j,j=1,2,3$.

Let us first consider the right hand side of Equation
(\ref{B.4eqa}) for $h$ replaced by $h_1$ We introduce the
variables ${\bf K}_j=\sum_{l=j}^n{\bf k}_l$ for $j=3,\ldots,n$ and
we set $\cos\theta_3={\bf K}_3 \cdot{\bf k}/(|{\bf K}_3||{\bf
k}|)$ and $\cos \theta_j={\bf K}_{j-1}\cdot {\bf K}_{j}/(|{\bf
K}_{j-1}||{\bf K}_j|)$ for $j=4,\ldots,n$. We then get
\begin{eqnarray*}
\rho({\bf k},{\bf k}_3,\ldots,{\bf k}_n)&=&\rho ({\bf k},(|{\bf
K}_3|,\cos\theta_3),\ldots,(|{\bf K}_n|, \cos \theta_n))\\
&=&(|{\bf k}|^2+|{\bf K}_3|^2+2|{\bf k}||{\bf K}_3|\cos \theta_3
+m^2)^{1/2}\\ &+&\sum_{l=3}^{n-1}(|{\bf K}_l|^2+|{\bf
K}_{l+1}|^2-2|{\bf K}_l||{\bf K}_{l+1}|\cos
\theta_{l+1}+m^2)^{1/2}\\ &+&(|{\bf K}_n|^2+m^2)^{1/2}.
\end{eqnarray*}

If we now change variables ${\bf k}_2,\ldots,{\bf k}_n\to {\bf
K}_3,\ldots,{\bf K}_n$ in (\ref{B.4eqa}) and we then pass over to
spherical coordinates $(|{\bf K}_l|,\cos \theta_l,\varpi_l),
l=3,\ldots,n$ where $(\cos \theta_l,\varpi_l)$ are coordinates on
the sphere $S^{d-2}$ (and the surface element on $S^{d-2}$ is
denoted by $d\cos\theta_ld\varpi_l$) we get
\begin{eqnarray}
\label{B.5eqa} &&\int_{(S^{d-2}\times (0,\infty))^{\times n-2}}
h_1(k,(|{\bf K}_3|,\cos\theta_3,\varpi_3),\ldots ,(|{\bf
K}_n|,\cos \theta_n,\varpi_n))\nonumber \\ &\times& \delta (\rho
({\bf k},(|{\bf K}_3|,\cos\theta_3),\ldots,(|{\bf K}_n|, \cos
\theta_n))+k^0)\nonumber \\ &\times& d\cos\theta_3d\varpi_3|{\bf
K}_3|^{d-2}d|{\bf K}_3|\cdots d\cos\theta_nd\varpi_n|{\bf
K}_n|^{d-2}d|{\bf K}_n|
\end{eqnarray}
where  we have written the function $h_1$ as a function of the new
variables.

Using the formula $$ \delta(\rho(x)-a)=\sum_{y:\rho(y)=a}{1\over
|\rho'(y)|}\delta(x-y) $$ which holds if $\rho'(y)\not =0$ if
$\rho(y)=a$ and setting
\begin{eqnarray*}
\varphi({\bf k},{\bf K}_3,\cos\theta_3)&=&{d\over
d\cos\theta_3}\rho ({\bf k},(|{\bf
K}_3|,\cos\theta_3),\ldots,(|{\bf K}_n|, \cos \theta_n))\\
&=&{|{\bf K}_3||{\bf k}|\over (|{\bf k}|^2+|{\bf K}_3|^2+2|{\bf
k}||{\bf K}_3|\cos \theta_3 +m^2)^{1/2}}
\end{eqnarray*}
we get for (\ref{B.5eqa})
\begin{eqnarray}
\label{B.6eqa} &&\int_{(S^{d-2}\times (0,\infty))^{\times n-2}}
h_1(k,(|{\bf K}_3|,\cos\theta_3,\varpi_3),\ldots ,(|{\bf
K}_n|,\cos \theta_n,\varpi_n))\nonumber \\ &\times& {\delta (\cos
\theta_3-\psi( k,|{\bf K}_3|,(|{\bf
K}_4|,\cos\theta_4),\ldots,(|{\bf K}_n|, \cos
\theta_n)))\over\varphi({\bf k},{\bf
K}_3,\cos\theta_3)}\nonumber\\&\times&d\cos\theta_3d\varpi_3|{\bf
K}_3|^{d-2}d|{\bf K}_3|\cdots d\cos\theta_nd\varpi_n|{\bf
K}_n|^{d-2}d|{\bf K}_n|
\end{eqnarray}
where
\begin{eqnarray*}
&&\psi( k,|{\bf K}_3|,(|{\bf K}_4|,\cos\theta_4),\ldots,(|{\bf
K}_n|, \cos \theta_n))\\ &=& \bigg[ \Big(
-k^0-\sum_{l=4}^{n-1}(|{\bf K}_l|^2+|{\bf K}_{l+1}|^2-2|{\bf
K}_l||{\bf K}_{l+1}|\cos \theta_{l+1}+m^2)^{1/2}\\ &-&\left.
(|{\bf K}_n|^2+m^2)^{1/2}\Big)^2-|{\bf k}|^2-|{\bf
K}_3|^2-m^2\bigg] \right/(2|{\bf k}||{\bf K}_3|)
\end{eqnarray*}
is a smooth function on the set of arguments which are in the
support of $h_1$. Furthermore, since $|{\bf k}|,|{\bf K}_3|>1$ in
the support of $h_1$, derivatives $(\partial^{|\alpha|}/\partial
k^\alpha)\psi$ $(\partial^{|\alpha|}/\partial k^\alpha)\cos
\theta$ also are bounded on the support of $h_1$ for any
multiindex $\alpha$.

We now set $\tilde h_1=h_1/\varphi$ and we get that $\tilde
h_1\in{ \cal S}_{1,2}(\R^{d+(d-1)(n-2)},\C)$ with $\| \tilde
h_1\|_{2,L}\leq c_L\|h_1\|_{2,L}$ for some $c_L>0$, $l\in\N_0$.

Consequently, we get for a
multinindex $\alpha$ with $|\alpha|=0,1,2$
\begin{eqnarray*}
&& \bigg| {\partial^{|\alpha|}\over\partial
k^\alpha}\int_0^\infty\int_{-1}^1\delta(\cos\theta_3-\psi )\tilde
h_1~d\cos \theta_3|{\bf K}_3|^{d-2}d|{\bf K}_3|\bigg|\\
&\leq&c\|\tilde h_1\|_{2,L'}{1\over
(1+|k|^2)^{L/2}}\prod_{l=4}^n{1\over (1+|{\bf K}_l|^2)^{d/2}},
\end{eqnarray*}
for $c$ sufficiently large and $L'=\max\{ L,d\}$. If we insert
this estimate into (\ref{B.6eqa}), we get $\|g_1\|_{2,L}\leq
c_L\|\tilde h_1\|_{2,L'}\leq c_L'\|h\|_{2,L'}$. If we can prove
similar estimates for $g_2,g_3$, the proof is finished.

This is simple for $g_2$: We consider $h_2$ and $g_2$ as functions
of the new variable ${\bf k}'={\bf k}+{\bf a}$ for some ${\bf
a}\in\R^{d-1}$ with $|{\bf a}|\geq 3$. Then function $h_2$ the in
these new variables fulfills the same conditions as $h_1$ before
and we get the desired estimate.

It remains to show the estimate for $g_3$. Let $k,{\bf
K}_3,\ldots,{\bf K}_n$ the coordinates introduced above. We define
the vector field ${\bf b}={\bf b}({\bf K}_n)=3{\bf K}_n/|{\bf
K}_n|$ and we introduce new variables ${\bf k}'={\bf k}-{\bf b}$,
${\bf K}'_l={\bf K}_l+{\bf b}$, $l=3,\ldots,n$. In the polar
coordinates $|{\bf k}'|,|{\bf K}_l'|,\cos \theta_3'={\bf
k}'\cdot{\bf K}_3'/(|{\bf k}'||{\bf K}_3'|),\cos \theta'_l={\bf
K}'_{l-1}\cdot {\bf K}'_{l}/(|{\bf K}'_{l-1}||{\bf K}'_{l}|)$ we
then get for $\rho(k,{\bf k}_3,\ldots,{\bf k}_n)$:
\begin{eqnarray*}
&&(|{\bf k}'|^2+|{\bf K}'_3|^2+2|{\bf k}'||{\bf K}_3'|\cos
\theta'_3 +m^2)^{1/2}\\ &+&\sum_{l=3}^{n-1}(|{\bf K}'_l|^2+|{\bf
K}'_{l+1}|^2-2|{\bf K}'_l||{\bf K}'_{l+1}|\cos
\theta_{l+1}'+m^2)^{1/2}\\ &+&((|{\bf K}'_n|+3)^2+m^2)^{1/2}
\end{eqnarray*}
and we can proceed as before, since $|{\bf k}'|,|{\bf K}_3'|>1$ on
the support of $h_3$.

\small

\noindent {\bf Acknowledgments.} We thank C. Becker, D. Buchholz,
S. Doplicher, R. Gielerak, O. W. Greenberg, K. Iwata, T. Kolsrud,
G. Morchio, F. Strocchi and J.-L. Wu for interesting discussions.
This work was made possible through financial support of D.F.G.
SFB 237 and the ``Hochschulsonderprogramm III'' of the federation
and lands of Germany via a D.A.A.D. sholarship for the second
named author.

\end{document}